\documentclass[twocolumn,superscriptaddress,showpacs]{revtex4}
\usepackage{amsmath}
\usepackage{graphicx}

\begin{document}

\title{Importance of on-site interaction in graphene.}

\author{Hari P. Dahal}
\affiliation{Department of Physics, Boston College, Chestnut Hill,
 MA, 02467}
 \affiliation{Theoretical Division, Los Alamos National Laboratory,
Los Alamos, New Maxico 87545}

\author{Jean-Pierre Julien}
\affiliation{ Institute Neel, CNRS, Associated to Unviersity J.
Fourier, B.P. 166, 38042 Grenoble Cedex 9, France}
 \affiliation{Theoretical Division, Los Alamos National Laboratory,
Los Alamos, New Maxico 87545}

\author{A. V.  Balatsky}
\affiliation{Theoretical Division, Los Alamos National Laboratory,
Los Alamos, New Maxico 87545} \affiliation{Center for Integrated
Nanotechnology, Los Alamos National Laboratory, Los Alamos, New
Maxico 87545}

\date{\today}

\begin{abstract}
We use the Gutzwiller method to investigate the importance of the
on-site Coulomb interaction in graphene. We apply it to Hubbard
Hamiltonian to study the renormalization of the kinetic energy in
graphene due to the on-site Coulomb interaction. We find that a
reasonable strength of the interaction has a very weak effect in
reducing the kinetic energy. Hence we predict that the
Brinkmann-Rice metal-insulator transition in graphene is not
possible. The effect is understood in terms of the high kinetic
energy in graphene.
\end{abstract}

\maketitle
Graphene is a two dimensional (2D) crystal of carbon atoms. It is
an allotropy form of Carbon in which the atoms form a honeycomb
lattice. The experimental success of creating the graphene
\cite{novoselov2005,novoselov22005} has renewed the interest of
studying the electronic properties of the two dimensional electron
system. It is because the 2D crystals were predicted to be
thermodynamically unstable at finite
temperature.\cite{peierls1935,landau1937} Graphene is stabilized
by strong covalent bonding between carbon atoms and a small
crumpling in the third dimension which increases the elastic
energy. It was known theoretically that the low energy excitations
of graphene behave as massless Dirac fermion \cite{wallace1947};
which leads to different electronic properties in this system
compared to that of the conventional two dimensional electron
system. In fact, the observation of the half integer quantum Hall
effect in graphene is one of the examples.\cite{novoselov22005}

In graphene the nearest neighbor hopping, which sets a scale to
the kinetic energy, is very high. In such case we need to
understand how important the role of the Coulomb interaction will
be in determining the electronic properties. Our previous study
shows that the effect of the long range Coulomb interaction is not
important in single layer graphene. We have shown that this system
can not have charge inhomogeneous states
(\cite{dahal2006,dahal2007}). It has also been shown that this
system can not undergo a ferromagnetic
transition(\cite{peres2005}). Even if there is a consensual
agrement about the fact that the Coulomb interaction does not have
a strong effect on the electronic properties of graphene, we do
here an explicit calculation to benchmark this statement. We study
the importance of the on-site Coulomb interaction using Hubbard
model in Gutzwiller approach.\cite{gutzwiller1963,gutzwiller1965}
In Gutzwiller frame work the Hubbard model leads to the
possibility of a metal insulator transition at half filling, which
is also known as the Brinkmann-Rice transition. The idea is to
look at the effect of the on-site Coulomb repulsion on the
renormalization factor of the kinetic energy which is a direct
measure of the correlation effect. In graphene we find that the
renormalization in the kinetic energy is very small. Hence we
infer that the system can not be even close to the Brinkmann-Rice
Metal-Insulator transition.

First, we introduce the Gutzwiller method briefly for which we
closely follow Ref. \cite{julien2006}. Among numerous theoretical
approaches, the Gutzwiller method provides a transparent physical
interpretation in terms of the atomic configurations of a given
site. Originally, it was applied to the one-band Hubbard model
Hamiltonian. \cite{hubbard1963} In graphene also the electronic
band close to the Fermi level is formed by the hybridization of
neighboring $p_z$ orbital. So introducing single Slater Koster
\cite{slaterkoster1954} parameter $pp\pi \sim 3.0 eV$ it is
straight forward to use the method in this system. We begin with
the Hubbard Hamiltonian,
\begin{equation}
H = H_{kin} + H_{int},
\label{hamiltonian}
\end{equation}
with,
\begin{equation}
H_{kin}=\sum_{i \neq j, \sigma} t_{ij}c_{i\sigma}^\dagger
c_{j\sigma},
\end{equation}
and,
\begin{equation}
H_{int}= U\sum_i n_{i\uparrow}n_{i\downarrow}.
\end{equation}
The Hamiltonian contains a kinetic part $H_{kin}$ with a hopping
integral $t_{ij}$ from site $j$ to $i$, and an interaction part
with a local Coulomb repulsion $U$ for electrons on the same site.
$c_{i\sigma}^\dagger$ ($c_{j\sigma}$) is the creation
(annihilation) operator of an electron at site $i$ with up or down
spin $\sigma$. $n_{i\sigma}=c_{i\sigma}^\dagger c_{i\sigma}$
measures the number (0 or 1) of electron at site $i$ with spin
$\sigma$. The Hamiltonian, Eq. \ref{hamiltonian}, contains the key
ingredients for correlated up and down spin electrons on a
lattice: the competition between delocalization of electrons by
hoping and their localization by the interaction. It is one of the
mostly used models to study the electronic correlations in solids.

In the absence of the interaction $U$, the ground state is
characterized by the Slater determinant comprising the
Hartree-like wave functions (HWF) of the uncorrelated electrons,
$|\psi_0\rangle$. When $U$ is turned on, the weight of the doubly
occupied sites will be reduced because they cost an additional
energy $U$ per site. Accordingly, the trial Gutzwiller wave
function (GWF) $|\psi_G\rangle$ is built from the  HWF
$|\psi_0\rangle$,
\begin{equation}
|\psi_G\rangle = g^D|\psi_0\rangle.
\end{equation}
The role of $g^D$ is to reduce the weight of the configurations
with doubly occupied sites, where $D=\sum_i
n_{i\uparrow}n_{i\downarrow}$ measures the number of double
occupations and $g (<1)$ is a variational parameter. In fact, this
method corrects the mean field (Hartree) approach for which up and
down spin electrons are independent, and, some how, overestimates
configurations with double occupied sites. Using the Rayleigh-Ritz
principle, this parameter is determined by minimization of the
energy in the Gutzwiller state $|\psi_G\rangle$, giving an upper
bound to the true unknown ground state energy of $H$. Note that to
enable this calculation to be tractable, it is necessary to use
the Gutzwiller's approximation which assumes that all
configurations in the HWF have the same weight.

Nozieres \cite{nozieres1986} proposed an alternative way which
shows that the Gutzwiller approach is equivalent to the
renormalization of the density matrix in the GWF. It can be
formalized as
\begin{equation}
\rho_G = T^\dagger \rho_0 T.
\label{renormalizeddensitymatrix}
\end{equation}
The density matrices $\rho_G=|\psi_G\rangle \langle \psi_G |$ and
$\rho_0 = |\psi_0 \rangle \langle \psi_0 |$ are projectors on the
GWF and HWF respectively. $T$ is an operator which is diagonal in
the configuration basis; $T=\Pi_i T_i$ where $T_i$ is a diagonal
operator acting on site i,
\begin{equation}
T_i|L_i,L' \rangle=\sqrt{\frac{p(L_i)}{p_0(L_i)}}|L_i,L' \rangle.
\label{tioperator}
\end{equation}
Here, $L_i$ is an atomic configuration of the site $i$, with
probability $p(L_i)$ in the GWF and $p_0(L_i)$ in the HWF
respectively, whereas $L'$ is a configuration of the remaining
sites of the lattice. Note that this prescription does not change
the phase of the wave function as the eigenvalues of the operators
$T_i$ are real. The correlations are local, and the configuration
probabilities for different sites are independent.

The expectation value of the Hamiltonian is given by,
\begin{equation}
\langle H \rangle_G = Tr(\rho_G H).
\label{hamiltonian_gutzwiller}
\end{equation}
The mean value of the on-site operators is exactly calculated with
the double occupancy probability, $d_i=\langle
n_{i\uparrow}n_{i\downarrow} \rangle_G$. $d_i$ is the new
variational parameters replacing $g$. Using Eqs.
(\ref{renormalizeddensitymatrix},\ref{tioperator}), the two-sites
operator contribution of the kinetic energy can be written as,
\begin{widetext}
\begin{equation}
\langle c_{i\sigma}^\dagger c_{j\sigma}\rangle_G = Tr(\rho_G
c_{i\sigma}^\dagger c_{j\sigma})=\langle c_{i\sigma}^\dagger
c_{j\sigma}
\rangle_0\sum_{L_{-\rho}}\sqrt{\frac{p(L'_\sigma,L_{-\sigma})}{p_0(L'_\sigma)}}
\sqrt{\frac{p(L_\sigma,L_{-\sigma})}{p_0(L_\sigma)}},
\end{equation}
\end{widetext}
where $L'_\sigma$ and $L_\sigma$ are the only two configurations
of spin $\sigma$ at sties $i$ and $j$ that give non-zero matrix
element to the operator in the brackets. The summation is
performed over the configurations of opposite spin $L_{-\sigma}$.
The probabilities $p_0$ in the HWF depend only on the number of
electrons, whereas the $p$ in the GWF also depends on $d_i$.

After some elementary algebra, one can show that the Gutzwiller
mean value can be factorized into,
\begin{equation}
\langle c_{i\sigma}^\dagger
c_{j\sigma}\rangle_G=\sqrt{q_{i\sigma}} \langle
c_{i\sigma}^\dagger c_{j\sigma}\rangle_0 \sqrt{q_{j\sigma}},
\label{normalizedenergy}
\end{equation}
where these renormalization factors $q_{i\sigma}$ are local and
can be expressed as,
\begin{equation}
\sqrt{q_{i\sigma}}=\frac{(\sqrt{1-n_{i\sigma}-n_{i-\sigma}+d_i}+\sqrt{d_i})
\sqrt{n_{i-\sigma}-d_i}}{\sqrt{n_{i\sigma}(1-n_{i\sigma})}}.
\end{equation}
We used $n_{i\sigma}$ as a shorthand notation for $\langle
n_{i\sigma} \rangle$, the average number of electrons on the
considered "orbital-spin" in the HWF, which could be site and/or
spin independent if the state is homogeneous and/or paramagnetic;
this is the case we consider here.

As seen in Eq. \ref{normalizedenergy}, the kinetic energy of the
non-interacting electrons state, $ \langle c_{i\sigma}^\dagger
c_{j\sigma}\rangle_0 = \varepsilon^0_{kin}$, is renormalized by a
factor of $q$ which is smaller than one in the correlated state,
and equal to one in the HWF. This factor can be interpreted as a
direct measure of the correlation effect. Indeed Vollhardt
\cite{vollhardt1984} has shown that $1/q=m^*/m$ where $m^*$ is the
effective mass and $m$ is the bare mass of the electron. Thus a
$q$ close to $1$ corresponds to less correlated electron system
and smaller $q$ value reflects enhancement of the correlation
effect. Eq. \ref{hamiltonian_gutzwiller} leads to the variational
energy,
\begin{equation}
E(d)=\langle H \rangle_G=2q \varepsilon^0_{kin} + Ud,
\end{equation}
which can be minimized with respect to the variational parameter
$d$. In the case of half filling (n=1/2), we can show that the
minimization condition is given by,
\begin{equation}
d =\frac{1}{4}(1-\frac{U}{16 \varepsilon^0_{kin}}),
\end{equation}
and,
\begin{equation}
q =(1-\frac{U^2}{(16 \varepsilon^0_{kin})^2}).
\end{equation}
It implies that if the Coulomb repulsion $U$ exceeds a critical
value $U_c = 16 \varepsilon^0_{kin}$, $q$ is equal to zero,
leading to an infinite quasiparticle mass with a Mott-Hubbard
Metal-Insulator transition which is also known as "the
Brinkmann-Rice transition" \cite{brinkmann1970}, as these authors
first applied the Gutzwiller approximation to the Metal-Insulator
transition.

To determine the magnitude of $U_c$ we need to calculate the
kinetic energy of the $p_z$-electrons in graphene. We use the
recursion method to first calculate the local density of states
(LDOS), i.e., projected on a given orbital:
\begin{equation}
\begin{split}
N_{{\psi}_1}(E)=\sum_n \delta (E-E_n)|\langle \psi_1 |n \rangle |^2 \\
= \frac{-1}{\pi}\lim_{\eta \rightarrow 0} \langle \psi_1
|\frac{1}{E+i\eta - H}|\psi_1 \rangle,
\end{split}
\end{equation}
where $|n \rangle$ are the eigenstates associated with eigenvalues
$E_n$, and using it we calculate the kinetic energy. Let us
briefly remind the recursion method of finding the density of
states. We follow Ref. \cite{haydoc1980} in describing the
recursion method.

For a given normalized state $|\psi_1 \rangle$ in which we want to
calculate the LDOS, we can always associate a recursion basis
which is constructed by a Schmidt orthogonalization procedure
starting from the set of states $|\psi_1\rangle$,
$H|\psi_1\rangle$,
$H^2|\psi_1\rangle$,...,$H^{N-1}|\psi_1\rangle....$. Let us
consider $H|\psi_1\rangle$. We can decompose it in a component
parallel to $|\psi_1\rangle$ and a component orthogonal to
$|\psi_1\rangle$. So we can write,
\begin{equation}
H|\psi_1\rangle=a_1|\psi_1\rangle + b_1 |\psi_2\rangle.
\end{equation}
If the coefficient $b_1$ is chosen real and positive then
$|\psi_2\rangle$ can be defined in a unique way. Consider then
$H|\psi_2\rangle$, this vector can be decomposed in a component
parallel to the space spanned by $|\psi_1\rangle$,
$|\psi_2\rangle$ and a component orthogonal to this space. Then we
get,
\begin{equation}
H|\psi_2\rangle=a_2|\psi_2\rangle+b'_1|\psi_1\rangle+b_2
|\psi_3\rangle.
\end{equation}
Furthermore, since $H$  is a hermitian operator we deduce that
$b'_1=b_1$. If the coefficient $b_2$ is chosen real and positive
then $|\psi_3\rangle$ can also be defined in a unique way. We can
repeat the process leading to the construction of a set of states
$|\psi_n\rangle$ which are orthonormal and satisfy,
\begin{equation}
H|\psi_n\rangle=a_n|\psi_n\rangle+b_{n-1}|\psi_{n-1}\rangle +
b_n|\psi_{n+1}\rangle.
\end{equation}

In the basis of $|\psi_n\rangle$ the Hamiltonian is tridiagonal.
An important property of the sates $|\psi_n\rangle$ is that they
spread progressively from the initial state. For example if
$|\psi_1\rangle$ is located on an atomic orbital $p_z$ then the
state $|\psi_n\rangle$ can have non-zero components only on
orbital that is reached by $n-1$ applications of the Hamiltonian
$H$, i.e., $n-1$ hopping from initial orbital.

The recursion basis establishes an order between states that are
attained at shorter or longer times when the initial states is
$|\psi_1\rangle$. Thus, roughly speaking one expects that it is
important to treat properly the states with smallest indices $n$,
and in general it will have smaller effect if states with larger
indices are approximated or simply not considered.

\begin {figure}
\includegraphics[width= 7.0 cm, height = 7.0cm,angle=-90]{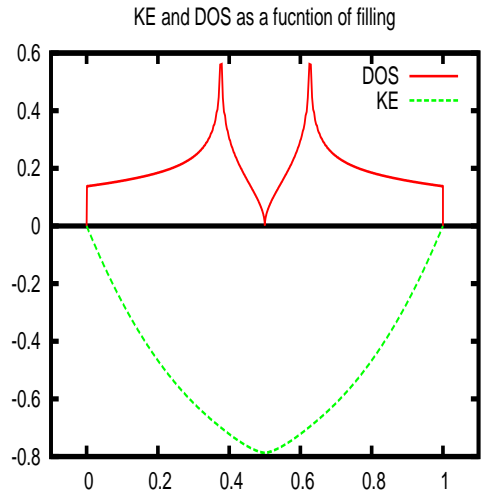}
\caption{(Color online) The density of states (DOS) and the
kinetic energy (KE) as a function of filling factor in graphene.
Energy is measured in unit of $t$.}
\end{figure}

From a technical point of view the interest of the recursion basis
is that it can be used to calculate vectors like
$f(H)|\psi_1\rangle$ where $f(H)$ is any operator that is a
function of the Hamiltonian $H$. This vector can be decomposed in
the basis of the states $|\psi_n\rangle$ as,
\begin{equation}
f(H)|\psi_1\rangle=\sum_n |\psi_n\rangle \langle \psi_n
|f(H)|\psi_1\rangle.
\end{equation}
When one applies this method to a function, $f(H)=(z-H)^{-1}$, one
can show that,
\begin{equation}
 R(z)=\langle \psi_1 | \frac{1}{z-H}|\psi_1 \rangle =
\frac{1}{z-a_1-\frac{b_1^2}{z-a_2-\frac{b_2^2}{z-a_3-.....}}}.
\end{equation}
In the simplest case where the spectrum does not have a gap, the
coefficients $a_n$ and $b_n$ tend to a limit $a_\infty$ and
$b_\infty$ respectively. If these limits are obtained to a good
precision one can write
\begin{equation}
R(z) =
\frac{1}{z-a_1-\frac{b_1^2}{z-a_2-\frac{b_2^2}{z-a_3-.....-\frac{b_{n-1}^2}{z-a_n-b_n^2R_T(z)}}}},
\end{equation}
where,
\begin{equation}
R_T(z)=\frac{1}{z-a-\frac{b^2}{z-a-\frac{b^2}{z-a-....}}},
\end{equation}
is known as the terminator of the series. This infinite expansion
means that $R_T(z)$ obeys a simple relation,
\begin{equation}
R_T(z)=\frac{1}{z-a-b^2R_T(z)},
\end{equation}
from which one deduces,
\begin{equation}
R_T(z)=\frac{z-a\pm \sqrt{(z-a)^2-4b^2}}{2b^2}.
\end{equation}
So $R_T(z)$ is the resolvent for a semi-elliptic density of states
with energies in the range $a-2b\leq E \leq a+2b$. In the case of
graphene, due to the the symmetry of the honeycomb lattice, all
$a_n =0$. On another hand the presence of the pseudo gap in the
middle of the band leads to two alternating limits for $b_{odd}$
and $b_{even}$. Thus we use a second order terminator, which can
be written as,
\begin{equation}
R_T(z)=\frac{1}{z-a_{\infty}-\frac{b_{\infty}^{2}}{z-
a_{\infty}'-{b'_{\infty}}^{2}R_T(z)}},
\end{equation}
which can be easily solved.

\begin {figure}
\includegraphics[width= 8.50 cm, height = 5.0cm]{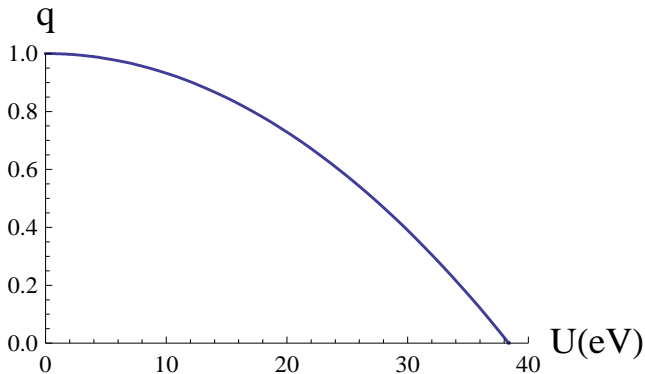}
\caption{ The kinetic energy renormalization factor as a function
of on-site Coulomb potential. For practical values of potential
the renormalization of the kinetic energy is very small.}
\end{figure}

This continued fraction expansion of the diagonal element of the
resolvent provides an efficient way of computing the LDOS using
the relation,
\begin{equation}
N(E)= \frac{-1}{\pi}ImR(E).
\end{equation}
Once we get the density of states as a function of filling factor,
we can calculate the kinetic energy per spin using,
\begin{equation}
\varepsilon_{kin}^0 = \int_{a-2b}^{\varepsilon_F} EN(E) dE.
\end{equation}

The calculated density of states and kinetic energy as a function
filling factor is presented in Fig. 1. The energy is measured in
terms of $t=3.0 eV$$(=pp\pi)$, the nearest neighbor hopping
energy. From the figure we can see that the kinetic energy per
spin at half filling, $\varepsilon_{kin}^0\cong 2.4 eV$. So the
critical on site repulsion required for the Brinkmann-Rice
transition is $U_C =  38.4 eV$ which is far from being a plausible
magnitude of the Coulomb interaction in graphene.

We also calculate the kinetic energy renormalization factor $q$ as
a function of the on site Coulomb energy. The result is shown in
Fig. 2. We can clearly see form the figure that the
renormalization of the kinetic energy due to the on-site Coulomb
interaction is very small for practical values of the interaction.
For example, even for the interaction as high as $10eV$ the
kinetic energy is reduced by only about $5 \%$.

In conclusion, we used Gutzwiller method to study the
Brinkmann-Rice transition in single layer graphene. We calculate
the density of states and the kinetic energy of electrons at half
filling using recursion method. We find that the kinetic energy of
electrons in graphene is very high such that the potential energy
required to have the metal insulator transition is unpractically
high. The on-site Coulomb interaction has very small effect on the
renormalization of the kinetic energy.

Acknowledgements: This work was carried out under the auspices of
the National Nuclear Security Administration of the U.S.
Department of Energy at Los Alamos National Laboratory under
Contract No. DE-AC52-06NA25396. Financial support from CNLS, and
Theoretical Division, T-11, as well as Delegation Generale Pour
L'Armement (DGA) under the contract 07.60.028.00.470.75.01 are
gratefully acknowledged. One of us (H.D.) is grateful to K. Bedell
and the Boston College for financial support.

\bibliographystyle{apsrev}
\bibliography{references}
\end{document}